\documentclass[aps,prl,twocolumn,groupedaddress,showpacs]{revtex4}
\usepackage{graphicx}
\usepackage{amssymb}

\begin{document}


\title{Projectile interactions in granular impact cratering}


\author{E.L. Nelson, H. Katsuragi\footnote{Current address: Department of Applied Science for Electronics and Materials, Kyushu University, Japan}, P. Mayor, and D.J. Durian}
\affiliation{
     Department of Physics \& Astronomy, University of
     Pennsylvania, Philadelphia, PA 19104, USA
}


\date{March 20, 2008; Revised: \today}

\begin{abstract}
We present evidence for the interactions between a ball and the container boundaries, as well as between two balls, that are mediated by the granular medium during impact cratering.  The presence of the bottom boundary affects the final penetration depth only for low drop heights with shallow filling, in which case, surprisingly, the penetration becomes deeper.  By contrast the presence of the side wall causes less penetration and also an effective repulsion.  Repulsion is also found for two balls dropped side-by-side.

\end{abstract}



\maketitle




The mechanical properties of close-packed non-cohesive granular media have both striking similarities and differences with those of ordinary solids and liquids~\cite{jnb, duran}.  This is reflected in their unusual resistance to penetration, a phenomenon utilized in many disciplines.  For example penetrometry is important for characterizing the nature of granular commodities, soils, and sediments.  It is similarly used for probing extraterrestrial surfaces \cite{ZarneckiN05}, as well as for designing the landing apparatus of spacecraft and the shape of anchors.  Penetration issues naturally arise for military ballistics and for planetary impact cratering.  Two experimental approaches have recently emerged to explore and isolate the key mechanisms that oppose granular penetration.  One is to measure the forces~\cite{schiffer04, schifferPRE04, KoehlerEL05, FranklinPRE06, SwinneyEPL07}, and the associated flows fields~\cite{BehringerPRL01, MeloEL06, LosertGM07}, required to achieve steady penetration at slow constant speed.  The other is to measure the kinematics and dynamics of projectile impact~\cite{jun, ciamarra, deBruyn2, detlef3, RPB04, HiroNP, DetlefPRL07, JaegerPRL07, Goldman08}.

An important result is that container boundaries can cause strong opposition to penetration, even at surprisingly large distances.  In particular, Refs.~\cite{schiffer04, schifferPRE04} demonstrated that the resistance force diverges exponentially as a slowly penetrating plate approaches bottom. This effect increases for larger plate diameters and thicknesses, for greater granular filling depths, and for smaller container diameters.  Container diameter effects were also considered in related penetration~\cite{KoehlerEL05, FranklinPRE06} and withdrawal~\cite{HorvathPRE96} experiments.  However, much less is known for the case of projectile impact.  Reduced granular filling depth was found to affect crater formation~\cite{BoudetPRL06}, but not the dynamics of deep penetration except for slight damped oscillations on approach to rest~\cite{HiroNP}.  Reduced container diameter was found to affect the penetration depth for sufficiently high impact speed~\cite{Goldman08}.  All these experiments suggest that boundaries may play a nontrivial and wider-than-suspected role in projectile impact.

In this paper we measure the change in the penetration of spherical projectiles dropped into a granular packing due to systematic changes in the filling depth of the medium as well as in the distance of the ball away from the container side wall.  In accord with prior observations, the presence of the side walls causes a shallower penetration -- but the change is smaller than anticipated.  The effect of the bottom wall is even smaller, and in the opposite direction.  In stark contrast to Refs.~\cite{schiffer04, schifferPRE04}, for deep filling the penetration is unchanged -- even for impacts that just reach the bottom -- and for shallow filling the penetration is actually deeper.  These seemingly contradictory results complicate the construction of a unified understanding and illustrate the counterintuitive nature of granular physics.


Here the granular medium consists of spherical glass beads with diameter range 250-350~$\mu$m, large enough that cohesion and interstitial air effects are unimportant~\cite{HiroNP}, and with bulk density $\rho_g=1.5$~g/cc.  The projectiles are steel spheres to which a narrow vertical acrylic rod is attached, so that both depth and lateral motion may be measured by either a microtelescope mounted on a height gauge and/or a high-speed video camera. The rod also facilitates in positioning and releasing the spheres, by use of metal tips and electromagnets and also by use of string.  Most data are for $D_b=1''$ diameter steel spheres, with total ball plus rod mass of 67~g.  Some data are for $1/2''$ and $2''$ diameter steel spheres with total ball plus rod masses of 11~g and 530~g, respectively.



\begin{figure}
\includegraphics[width=3.125in]{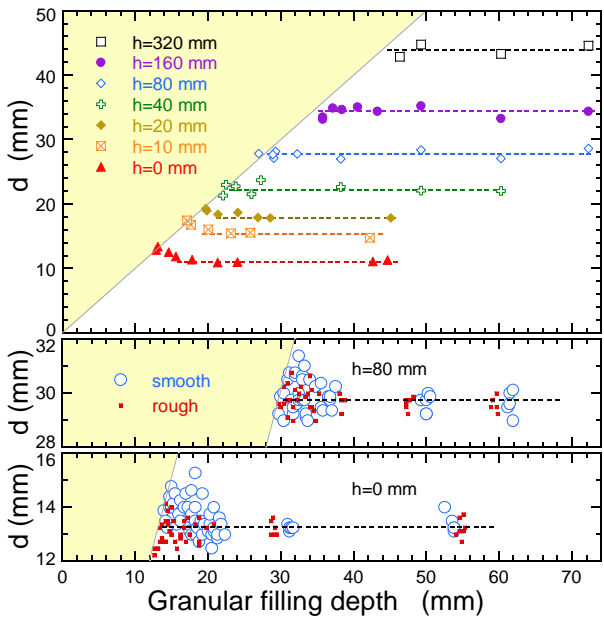}
\caption{(Color online) Penetration depth $d$ of a $1''$ diameter steel sphere vs the filling depth of the granular medium, for various free-fall drop heights $h$ as labeled.  The bottom two plots contrast results for smooth and rough boundaries for $h=80$~mm and $h=0$ respectively.  All symbols represent individual trials, except for the three smallest drop height in the top plot where the symbols represent the average of four trials.  The deep-medium asymptotes, where penetration is independent of filling depth, are indicated by horizontal dashed lines.  The forbidden regions of greater penetration than filling depth are shaded light yellow. For large enough drop heights that the ball strikes bottom with nonzero speed, the ball comes to rest without rebound at less than one grain diameter from bottom (not displayed).
\label{sanddepth}}
\end{figure}

The effects of the bottom boundary may be explored by variation of the granular filling depth.  For this, a controlled volume of medium is placed into a glass beaker with an inner diameter of either 10 or 12~cm.  The beaker is gently swirled by hand to produce an almost uniformly level surface and a random-close packing density of about 64\%~\cite{jun}.  The $1''$ diameter projectile is positioned, held from the rod, and released, by electromagnet.  The granular filling depth, the free-fall height $h$ of the projectile, and the final penetration depth $d$ of the projectile, are all measured by long-range height gauge.  Since $h$ is the distance between the top of the medium and the bottom of the ball, the impact speed is $\sqrt{2gh}$.  Results for penetration vs filling depth are shown in Fig.~\ref{sanddepth}, with runs taken at seven different free-fall drop heights ranging from $h=0$ to 320~mm.  As expected, a well-defined constant penetration depth is achieved for deep enough filling.  Surprisingly, however, no change in penetration is found for large drop heights -- even as the filling is decreased to the point that the projectile reaches the bottom of the container!  Shallow filling affects penetration only for small drop heights.  In such cases the penetration is actually greater than in the deep-filling limit.  Evidently the stopping force is smaller near the bottom wall, opposite to the steady penetration results~\cite{schiffer04}.  Perhaps the fast motion of the projectile, and its spherical shape, prevent excessive loading of force chains.  And perhaps the increased penetration is related to the slight decrease in penetration force that occurs for smooth-bottom cells just prior to the exponential increase~\cite{schifferPRE04}, though this would be expected for deep-filling too.

The observed effects of granular filling depth on penetration may be understood qualitatively as follows.  One contribution to the total stopping force is a rate-independent friction term $kz$ that depends on depth $z$ through the hydrostatic loading of grain-grain contacts~\cite{schiffer01, detlef3, LevPG, HiroNP}.  Since the local hydrostatic pressure depends only on the weight of medium above, this term is unaffected by shallow filling if the projectile motion does not load the grain-grain contacts.  Another contribution to the total stopping force is an inertial drag that arises from the transfer of momentum to a projectile-sized volume of the granular medium~\cite{LevPG, HiroNP}.  Since ordinarily the projectile sets the only length scale, this term should be proportional to $\rho_g {D_b}^2 v^2$.  However, if the distance $s$ between the projectile and the bottom of the container is less than $D_b$, then a smaller volume of medium is set into motion and the reduced inertial drag should scale more like $\rho_g D_b s v^2$.  This will cause greater penetration, particularly when the filling depth is less than $D_b$ so that the reduction in inertial drag is important even at initial impact when the $kz$ term is negligible.  For deep enough filling, by contrast, there should be little increase in penetration because the reduction in inertial drag does not occur until toward the end of impact, where $kz$ dominates the total stopping force. This picture is consistent with the main plot in Fig.~\ref{sanddepth}, where the penetration increases only for filling depths and penetrations less than about $D_b$.  It is also consistent with the bottom two plots in Fig.~\ref{sanddepth}, where 0.35~mm roughness (50 grit adhesive sandpaper) along the bottom boundary has no influence for $h=80$~mm but eliminates the attraction for $h=0$.


\begin{figure}
\includegraphics[width=3.125in]{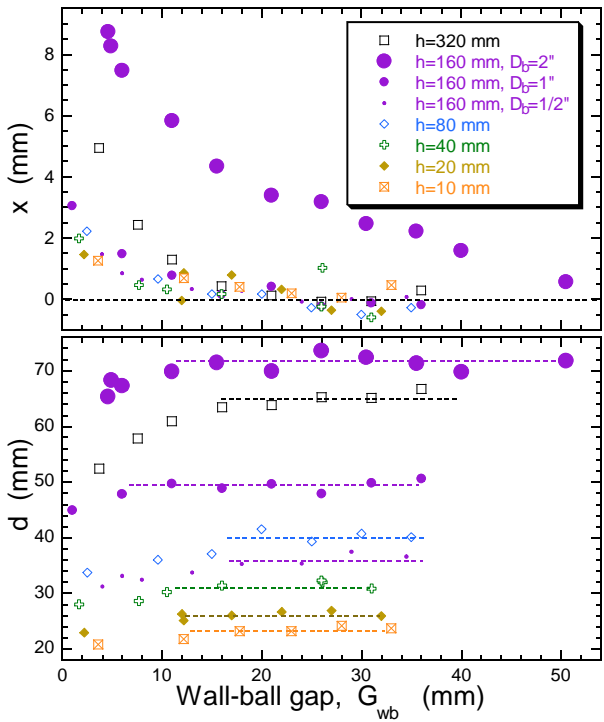}
\caption{(Color online) Horizontal displacement $x$ away from vertical side wall, and penetration depth $d$, vs initial gap $G_{wb}$ between wall and ball, for various free-fall drop heights $h$ and ball diameters $D_b$ (all are $1''$ except as labeled).\label{wallgap}}
\end{figure}

\begin{figure}
\includegraphics[width=3.125in]{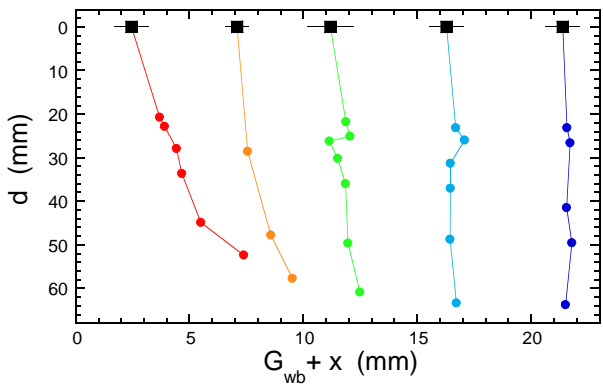}
\caption{(Color online) Penetration depth $d$ vs final distance from the side wall. The latter is given by the sum of initial wall-ball gap $G_{wb}$ plus the total horizontal displacement $x$ during impact.  Results are grouped together by color and connecting lines for different drop heights at similar values of the initial wall-ball gap, which is indicated by the solid black square and horizontal error bar.  The y-axis is inverted, since depth is measure downwards.  The sequence of free-fall drop heights are $h=\{10,20,40,80,160,320~{\rm mm}\}$ with the following exceptions: for $G_{wb}\approx 7$~mm there are no points for $h=\{10,20,80~{\rm mm}\}$; for $G_{wb}\approx 11$~mm there is a second point for $h=20$~mm; for $G_{wb}\approx 21$~mm there is no point for $h=40$~mm. \label{traj}}
\end{figure}

The second set of experiments is to consider the effects of one vertical boundary of the sample container, i.e.\ to consider the change in penetration when the initial gap $G_{wb}$ between wall and ball is small.  From the above discussion, one might expect reduced inertial drag on the side of the projectile near the wall; this would lead to deeper penetration and also to lateral motion of the projectile toward the wall.  To explore this, steel balls are dropped as above but now into a post- gas-fluidized granular medium in the same 19~cm inner-diameter fluidization cell used in Ref.~\cite{HiroNP}.  This sample preparation protocol gives a smooth level surface and a random loose packing fraction of about 59.5\%~\cite{ojha}. The final penetration depth $d$, and the net change $x$ in the gap between wall and ball, are extracted by video analysis of the location and angle of acrylic rod.  Results are plotted vs initial gap in Fig.~\ref{wallgap}.  For large gaps there is no effect. But when $G_{bw}$ is small, $x$ becomes positive and also the penetration becomes smaller; therefore, the effect of the side wall is to repel the projectile and also to increase the upward stopping force.  The same trends are also observed for the previous preparation protocol, where the granular medium is gently swirled to a smooth level state with higher density.

At present we have little explanation for the observed influence of a sidewall.  To illustrate it more graphically, the final penetration $d$ is plotted vs the the final wall-ball gap $G_{wb}+x$ in Fig.~\ref{traj}.  Data points with the same initial gap $G_{wb}$ are connected by line segments, so that the locus of stopping points appears like an impact trajectory that slows down and curves away from the side wall to an extent that depends on initial proximity.  These faux ``trajectories'' serve to emphasize that the repulsion from the walls is stronger than the decrease in the upward stopping force.  They also emphasize that the effect is stronger for higher drop heights, deeper into the sample, where friction is stronger and inertia is smaller.  Hence we speculate that the motion of the projectile serves to load force chains between ball and wall, which causes the repulsion.  Note also in Fig.~\ref{traj} that there is essentially no repulsion for gaps greater than about one ball radius even for impacts as deep as five ball radii.  This is smaller than the effects found in Refs.~\cite{schiffer04, schifferPRE04, Goldman08}.  It is also smaller than for simple fluids, where a sedimenting ball migrates to the center of a cylinder no matter how large the radius~\cite{JosephJFM94}.


\begin{figure}
\includegraphics[width=3.125in]{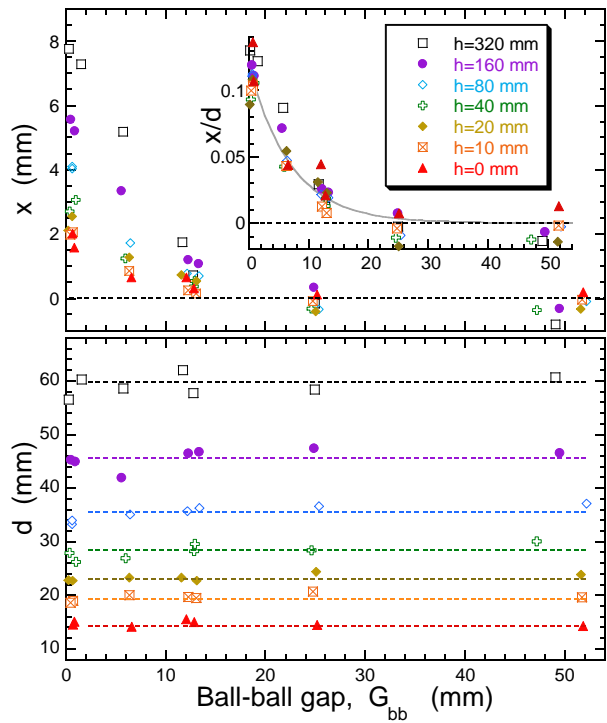}
\caption{(Color online) Horizontal displacement and penetration depth for two balls dropped side-by-side vs their initial gap, $G_{bb}$, for various free-fall drop heights as labeled.  The inset in the top plot shows horizontal displacement scaled by penetration depth, along with the fitting result $x/d=0.12\exp[-G_{bb}/(7~{\rm mm})]$ as a solid grey curve.\label{ballgap}}
\end{figure}

The final set of experiments is to consider the effects of a second projectile dropped side-by-side with the first, i.e.\ to consider the change in penetration when the initial gap $G_{bb}$ between two identical balls is small.  Based on results for the ball-wall experiments, one might expect two balls to repel each other and to penetrate less deeply.  To explore this, $1''$ steel balls are suspended by a light thread tied to the tops of the acrylic rods and draped over two horizontal $1/2''$ diameter posts fixed at variable separation.  The thread is burned with a butane lighter at midpoint, half way between the two posts, in order to achieve simultaneous impact, which is verified by high-speed video to within 2~ms.  As for the ball-wall experiments, the medium is prepared by post- gas-fluidization and the penetration depths $d$ and the change $x$ in ball-ball separation are found by analysis of the final rod positions.  Results are plotted in Fig.~\ref{ballgap} vs initial ball-ball gap $G_{bb}$ for a variety of drop heights.  As expected the two balls move apart, indicative of an effective repulsive interaction mediated by the grains.  Surprisingly, however, there is no detectable effect on the final penetration depths, which are found to depend only on drop height -- independent of the initial gap $G_{bb}$.  The ball-ball interaction is thus simpler than the ball-wall interaction.  Indeed, as illustrated by the inset, all data for change in separation can be made to collapse if scaled by the total penetration depth.  This is captured well by the empirical form $x/d = 0.12 \exp[ -G_{bb} / (7~{\rm mm}) ]$, shown as a solid gray curve.  We have no explanation for this in terms of the actual depth- and velocity-dependent forces on the projectiles.  Perhaps force chains between the two balls are loaded by their motion, similar to our intuition for the ball-wall effect.  An alternative speculation would be a greater shear rate, and hence a greater granular temperature and pressure, between ball-wall and ball-ball.

In conclusion it is simpler to avoid finite container-size effects than had been expected from prior penetration and impact experiments.  The filling depth, the distance from a side wall, and the distance from another projectile, must be larger than only the ball diameter.  This contrasts with the case of simple liquids, where the relevant length scale would be set by penetration depth.  Furthermore, the effect of the bottom boundary is opposite in sign to steady penetration results~\cite{schiffer04, schifferPRE04}.  The observed finite-size effects on granular impact may perhaps be understood in terms of changes in inertial drag and in the loading of force chains due to projectile motion.  However, other possible effects should not be discounted, such as slippage of rigid force chains along a smooth boundary or fluidization from a reflected shock wave.  Sorting out and quantifying these effects, and reconciling with the steady penetration experiments, call for further investigation \cite{Bertho}.

\begin{acknowledgments}
We than P.E.\ Arratia, S.R.\ Nagel, P.E.\ Schiffer, and P.B.\ Umbanhowar for helpful conversations.  This work was supported by the NSF through grants DMR-0704147 (DJD) and DMR06-48953 (ELN - Research Experience for Undergraduates), and also by the Japanese Society for the Promotion of Science (HK).\end{acknowledgments}

\bibliography{CraterRefs}

\end{document}